\documentclass[%
 reprint,
 amsmath,amssymb,longbibliography,aps,prx,]{revtex4-2}
\pdfoutput=1
\usepackage{graphicx,multirow}
\usepackage{dcolumn}
\usepackage{bm}
\usepackage{xcolor,colortbl}

\definecolor{d4blue}{rgb}{  0,  0.4470,    0.7410}
\definecolor{d12orange}{rgb}{0.8500    0.3250    0.0980}
\definecolor{g8yellow}{rgb}{0.    0.6    0.298}
\definecolor{ppurple}{rgb}{0.4940    0.1840    0.5560}
\newcommand{\mc}[2]{\multicolumn{#1}{c}{#2}}
\newcolumntype{k}{>{\columncolor{blue!20}}c}
\newcolumntype{r}{>{\columncolor{blue!10}}c}
\newcolumntype{d}{>{\columncolor{red!10}}c}

\newcommand{\down}{\downarrow}
\newcommand{\up}{\uparrow}
\newcommand{\spin}{\sigma}

\newcommand{\kv}{{\bf k}}
\newcommand{\qv}{{\bf q}}
\newcommand{\Qv}{{\bf Q}}

\usepackage{hyperref}

\begin{document}

\preprint{APS/123-QED}

\title{Pairing in the Two-Dimensional Hubbard Model from Weak to Strong Coupling}

\author{Astrid T. R\o mer,$^1$ Thomas A. Maier,$^2$ Andreas Kreisel,$^3$ Ilya Eremin,$^4$ P. J. Hirschfeld,$^5$ Brian M. Andersen$^1$}
\affiliation{
$^1$Niels Bohr Institute, University of Copenhagen, Lyngbyvej 2, DK-2100 Copenhagen,
Denmark\\
$^2$Center for Nanophase Materials Sciences, Oak Ridge National Laboratory, Oak Ridge, Tennessee 37831, USA\\
$^3$Institut f\"ur Theoretische Physik Universit\"at Leipzig
	D-04103 Leipzig, Germany\\
$^4$Institut f\"ur Theoretische Physik III, Ruhr-Universit\"at Bochum, D-44801 Bochum, Germany\\
$^5$Department of Physics, University of Florida, Gainesville, Florida 32611, USA
}%

\date{\today}

\begin{abstract}
The Hubbard model is the simplest model that is believed to exhibit superconductivity arising from purely repulsive interactions, and has been extensively applied to explore a variety of unconventional superconducting systems.  Here we study the evolution of the leading superconducting instabilities of the single-orbital Hubbard model on a two-dimensional square lattice as a function of  onsite Coulomb repulsion $U$  and band filling by calculating the irreducible particle-particle scattering vertex obtained from dynamical cluster approximation (DCA) calculations, and compare the results to both perturbative Kohn-Luttinger (KL) theory as well as the widely used random phase approximation (RPA) spin-fluctuation pairing scheme. Near half-filling we find remarkable agreement of the hierarchy of the leading pairing states between these three methods, implying adiabatic continuity between weak- and strong-coupling pairing solutions of the Hubbard model. The $d_{x^2-y^2}$-wave instability is robust to increasing $U$ near half-filling as expected. Away from half filling, the predictions of KL and RPA at small $U$ for transitions to other pair states agree with DCA at intermediate $U$ as well as recent diagrammatic Monte Carlo calculations. RPA results fail only in the very dilute limit, where it yields a $d_{xy}$ ground state instead of a $p$-wave state established by diagrammatic Monte Carlo and low-order perturbative methods, as well as our DCA calculations.  We discuss the origins of this discrepancy, highlighting the crucial role of the vertex corrections  neglected in the RPA approach. Overall, comparison
of the various methods over the entire phase diagram strongly suggests a smooth crossover of the superconducting interaction generated by local Hubbard interactions between weak and strong coupling.


\end{abstract}

\maketitle


\section{\label{sec:intro} Introduction}

Since the theoretical proposal by Kohn and Luttinger (KL) \cite{Kohn1965,Maiti2013}
that superconductivity can arise from purely repulsive electron interactions and the subsequent discovery of superconductivity in materials like heavy fermions, cuprates, organic Bechgaard salts, and iron-based superconductors, 
superconducting instabilities in models of interacting fermions have been extensively studied.  The Hubbard model\cite{Hubbard1963} has played an exceptional paradigmatic role in this discussion.  It is the simplest model of  fermions with local interactions, and was argued furthermore to be the appropriate effective model to describe unconventional superconductivity in correlated electron systems, notably in cuprates\cite{PWAnderson}. 
The  model has also been popular  because
the physics of pairing by spin fluctuations, originally suggested by Berk and Schrieffer~\cite{berk,Fay} in continuum models and extended by Scalapino and others to lattice Hubbard-type models\cite{Cyrot1986,Scalapino1986,Varma1986}, is rather simple to capture within the straightforward random phase approximation (RPA)\cite{Scalapino1999}. At present, theoretical studies of the Hubbard model constitute a growing research area as seen for example by several recent extensive comparisons between various state-of-the-art numerical methods providing updated benchmarks on e.g. the ground state energy, the self-energy, and competing order in the Hubbard model\cite{Gull_Simonscollaboration,Zheng1155}. An important next step is to compare and quantify the superconducting pairing instabilities within this model\cite{Maier2007}.

Close to half band filling, the 2D Hubbard model on a square lattice is known to exhibit strong $d$-wave pair correlations\cite{Scalapino_RMP}.
While a rigorous proof that $d$-wave superconductivity exists in the model at $T=0$ is lacking, the preponderance of  evidence from numerical  calculations\cite{doug,Bickers1989,Plekhanov2005,Paramekanti2004,Moriya2003,Tremblay2005,Aichhorn2006,Tremblay2012,Gull,deng15,staar,Simkovic_2016,zheng}, as well as weak and intermediate coupling renormalization group studies\cite{Zanchi96,Honerkamp2001,Honerkamp2001a,Zhai2009,Raghu2010,Halboth2000,Raghu2012,Eberlein2014}, strongly support this conclusion.  Rather less is known with high confidence at larger interaction $U$,  further away from the half-filled state, or with regard to subleading pair channels throughout the phase diagram.  It is convenient to  study the latter two questions using controlled perturbative methods or via RPA due to physical transparency and ease of implementation. Several authors, including the current ones, have applied various weak-coupling schemes to map out the leading superconducting instabilities as a function of e.g. doping and band parameters, displaying a rich mosaic of different pairing states\cite{Hlubina1999,Kondo2001,Guinea2004,Romer2015,Kreisel2017_JSNM,Arita2000,Yanagisawa2008}. Predictions of these studies for leading pairing instabilities throughout the phase diagram  appear to agree rather well  with recent diagrammatic Monte Carlo calculations that should be well-controlled and able to reach somewhat higher  $U$\cite{deng15}. However, the general question of how the pairing in the 2D Hubbard model changes as correlations are increased is still open.

In the special case close to half-filling, the Hubbard model reduces to the $t-J$ model as $U\rightarrow \infty$, and $t-J$ studies have also found $d$-wave pairing in this doping regime~\cite{RHA1987,KotliarLiu1988,Sorella2002,Prelovsek2005,Maier_PRL_2008,KotliarHaule2007,Spanu2008}.  This suggests the intriguing conclusion that the physics of pairing at strong coupling is similar to, or at least evolves continuously from, that at weak coupling. However, away from half-filling little is known about this crossover. With this question in mind, we calculate the superconducting pairing vertex of the Hubbard model via numerical solutions of the Bethe-Salpeter equation obtained in the dynamical cluster approximation (DCA) with quantum Monte Carlo (QMC) impurity solver.  This approximation\cite{maier05,gull11} is known to provide an accurate estimate of the pairing vertex over a range of intermediate strength $U$ values appropriate for the cuprates\cite{maier06}. We then compare these results with both perturbative Kohn-Luttinger (KL) theory and the RPA scheme. The latter breaks down at higher $U$ due to the well-known magnetic instability inherent to the approximation, but is thought to work well at smaller interaction strengths. We find that the hierarchy of pairing eigenvalues of the linearized gap equation match up well between the methods in the region where they can be compared, providing further evidence that the pairing evolution is smooth from weak to strong coupling.  Results from the simple RPA agree spectacularly well with diagrammatic Monte Carlo\cite{deng15} over the entire doping range except at the smallest doping, where $p$-wave spin triplet pairing is stable over a much narrower region in the RPA than obtained in asymptotically exact results\cite{Chubukov92,Chubukov93}.  We discuss the reasons for this discrepancy.

\section{\label{sec:model} Model and methods}
We study the single-orbital Hubbard model defined on a 2D square lattice
\begin{eqnarray}
H&=&-\sum_{i,j, \sigma}t_{i,j}c_{i\sigma}^{\dagger}c_{j\sigma}+ \sum_{i \sigma} U  n_{i\sigma} n_{i\bar\sigma} - \sum_{i \sigma} \mu  n_{i\sigma},   
\label{eq:Hhub}
\end{eqnarray}
where $c_{i\spin}^\dagger /c_{i\spin}$ creates/annihilates an electron at lattice site $i$ with spin $\spin$, and $n_{i\spin}=c_{i\spin}^\dagger c_{i\spin}$ is the number operator of electrons with spin $\spin$ at site $i$.
The nearest-neighbor hopping sets the energy unit, $t=1$, and we include also next-nearest neighbor hopping $t'$.
The superconducting pairing originates from the repulsive Coulomb interaction and is treated numerically by three different approaches. First, we calculate
the full energy-resolved pairing kernel with inclusion of self-energy corrections, and solve the Bethe-Salpeter equation with Green's functions and irreducible particle-particle vertex obtained from Quantum Monte Carlo simulations using the dynamic cluster approximation (DCA). The details are explained in Sec.~\ref{sec:dca} below.
Second, we apply perturbative KL-theory, and lastly compare with the RPA spin-fluctuation method for pairing, both detailed in Sec.~\ref{sec:rpa}. In KL-theory only the second order diagrams enter the pairing theory, whereas in the RPA approach, the effective pairing interaction is evaluated diagrammatically by a selected class of diagrams that highlights the physics of nesting and pronounced spin fluctuations. Whereas KL theory is a controlled weak-coupling approach valid at small interactions $U$ (compared to the bandwidth), the solution of the Bethe-Salpeter equation by DCA is not restricted to a certain regime of Hubbard-$U$. However, this method is significantly heavier computationally and suffers from the sign problem~\cite{maier05}. This imposes constraints on the smallness of $U$ as well as the size of $t'$, doping, and cluster size\cite{maier05,gull11}. 

\subsection{\label{sec:BS} Pairing within the Dynamical Cluster Approximation}
\label{sec:dca}
For the Quantum Monte Carlo calculations, we use a dynamic cluster
approximation \cite{maier05} with a continuous-time auxiliary field
(CT-AUX) QMC solver \cite{gull11}. The DCA represents the bulk lattice by a
finite size cluster and uses coarse-graining of the reciprocal space to retain
information about the remaining bulk degrees of freedom. Within this cluster approach, the first Brillouin zone is divided into $N_c$ patches ${\cal P}_{\bf
K}$, each of which is represented by a cluster momentum $\bf K$, and within
which the self-energy $\Sigma({\bf k},i\omega_n)$ is assumed to be constant
and given by the cluster self-energy $\Sigma_c({\bf K},i\omega_n)$. One then
averages the Green's function over the patches ${\cal P}_{\bf K}$ to determine
the coarse-grained Green's function
\begin{equation} {\bar G}({\bf K},i\omega_n)=\frac{N_c}{N}\sum_{\bf k\in {\cal
P}_{\bf K}}[i\omega_n+\mu-\varepsilon_{\bf k}-\Sigma_c({\bf
K},i\omega_n)]^{-1}\,.
\end{equation} Here the sum is restricted to the $N_c/N$ momenta within the
patch about the cluster momentum ${\bf K}$. The corresponding bare propagator
${\cal G}_0({\bf K},i\omega_n) = [{\bar G}^{-1}({\bf
K},i\omega_n)+\Sigma_c({\bf K},i\omega_n)]^{-1}$ is then used together with
the interaction $U$ to set up the effective cluster problem, in which the
self-energy $\Sigma_c({\bf K},i\omega_n) = {\cal F}[{\cal G}_0({\bf
K},i\omega_n),U]$ is calculated with the CT-AUX QMC solver. This calculation
is repeated iteratively until the self-energy has converged. For further details, the
reader is referred to Ref.~\cite{maier05}.

After convergence, the two-particle Green's function in the particle-particle
channel with zero center of mass momentum and energy, $G_{c,2}(K,K') =
G^{\uparrow\downarrow\downarrow\uparrow}_{c,2}(K,-K,-K',K')$ is  calculated
for the cluster problem \cite{maier06}. Here $K=({\bf K},i\omega_n)$ and
$K'=({\bf K'},i\omega_{n'})$. The irreducible particle-particle vertex
$\Gamma^{pp}(K,K')$ is then extracted from the Bethe-Salpether equation
\begin{eqnarray} G_{c,2}(K,K') = {\bar G}(K){\bar G}(-K)\delta_{K,K'} +\nonumber \\
 \frac{T}{N_c}\sum_{K''}{\bar G}(K){\bar
G}(-K)\Gamma^{pp}(K,K'')G_{c,2}(K'',K'),
\end{eqnarray} and used in the DCA gap equation for the bulk lattice
\begin{equation}
-\frac{T}{N_c}\sum_{K'}\Gamma^{pp}(K,K'){\bar{\chi}}_0^{pp}(K')\phi_\alpha(K') =
\lambda_\alpha\phi_\alpha(K),
\label{eq:DCAgapeqn}
\end{equation} where the pairing kernel $G({\bf k},i\omega_n)G({\bf
-k},-i\omega_n)$ has been coarse-grained over the momenta of the DCA patches
${\cal P}_{\bf K}$ to give ${\bar
\chi}_0^{pp}(K) = N_c/N\sum_{{\bf k}\in {\rm P}_{\bf K}} G({\bf
k},i\omega_n)G({\bf -k},-i\omega_n)$. The solution of this eigenvalue equation
 gives the DCA results for the leading eigenvalues $\lambda_\alpha$ and
corresponding eigenvectors $\phi_\alpha(K)$.
We use a cluster size of $N=64$ for $U=2$ and $N=32$ for $U=4,6,8$ and temperature is set to $T=0.025,0.05,0.15,0.2$ for $U=2,4,6,8$, respectively.

\subsection{\label{sec:rpa} Pairing within Kohn-Luttinger and RPA spin-fluctuation theory}

\begin{figure}[tb]
\centering
  	  \includegraphics[angle=0,width=\linewidth]{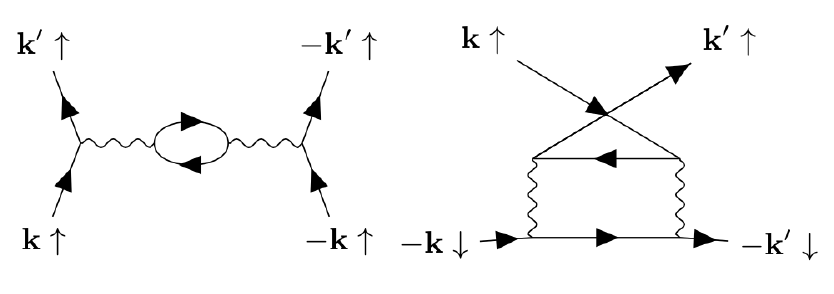}
\caption{Second order screening (bubble) and exchange (ladder) diagrams. Note that each interaction line $U$ depicted by a wiggly line connects opposite spins only. The bubble diagram contributes to same spin triplet pairing only, and singlet and opposite spin-triplet solutions arise from the ladder diagram after symmetrization and antisymmetrization, respectively, as stated in Eq.~(\ref{eq:KL}).}
\label{fig:KL}
\end{figure}

In both weak-coupling KL theory as well as in RPA spin-fluctuation mediated superconductivity, one derives an effective Cooper pair term of the form  
\begin{align}
 H_{\rm int}&=\frac{1}{2}\sum_{\kv,\kv'}V(\kv,\kv')c_{\kv'\up}^\dagger c_{-\kv'\down}^\dagger c_{-\kv\down} c_{\kv\up} 
 + \mbox{H.c.},
 \label{eq:Hrpa}
 \end{align}
 with $V(\kv,\kv')$ denoting the effective pairing vertex. In the KL approach, the vertex is obtained to second order in $U$. Since Hubbard interactions connect propagators of opposite spin only, this amounts to an evaluation of the diagrams depicted in Fig.~\ref{fig:KL}. Thus, the effective interaction is given by
 \begin{equation}
V_{KL}(\kv,\kv')=\frac{U^2}{2}[\chi_{0}(\kv+\kv')\pm\chi_{0}(\kv-\kv')],
\label{eq:KL}
    \end{equation}
where the upper (lower) sign corresponds to the singlet (triplet) channel and the bare spin susceptibility is given by the Lindhard function evaluated at zero energy
 $
  \chi_0(\qv)=\frac{1}{N}\sum_\kv \frac{f(\xi_{\kv+\qv})-f(\xi_{\kv})}{\xi_{\kv}-\xi_{\kv+\qv}}$, with $\xi_\kv=-2t(\cos(k_x)+\cos(k_y))-4t^\prime\cos(k_x)\cos(k_y)-\mu$. 

Within RPA, the screening (bubble) and exchange (ladder) diagrams depicted in Fig.~\ref{fig:KL} are summed to infinite order in $U$.
The bubble diagrams correspond to effective interactions through longitudinal fluctuations, while the ladder diagrams are due to exchange interactions mediated by transverse fluctuations. 
The final interaction between opposite spin electrons is
 \begin{align}
V(\kv,\kv')&= U+V_{\rm lo}^{\rm RPA}(\kv-\kv')+ V_{\rm tr}^{\rm RPA}(\kv+\kv'), \\
  V_{\rm lo}^{\rm RPA}(\kv-\kv')&= \frac{U^2}{2}\Big[\frac{\chi_0(\kv-\kv')}{1-U\chi_0(\kv-\kv')}-\frac{\chi_0(\kv-\kv')}{1+U\chi_0(\kv-\kv')}\Big], \nonumber \\
    V_{\rm tr}^{\rm RPA}(\kv+\kv')&= \frac{U^2\chi_0(\kv+\kv')}{1-U\chi_0(\kv+\kv')}.
    \end{align}
In this study, we restrict ourselves to the paramagnetic phase, where the longitudinal and transverse spin susceptibilities are the same and all triplet channels are degenerate.

 \begin{figure}[t!]
\centering
  	  \includegraphics[angle=0,width=\linewidth]{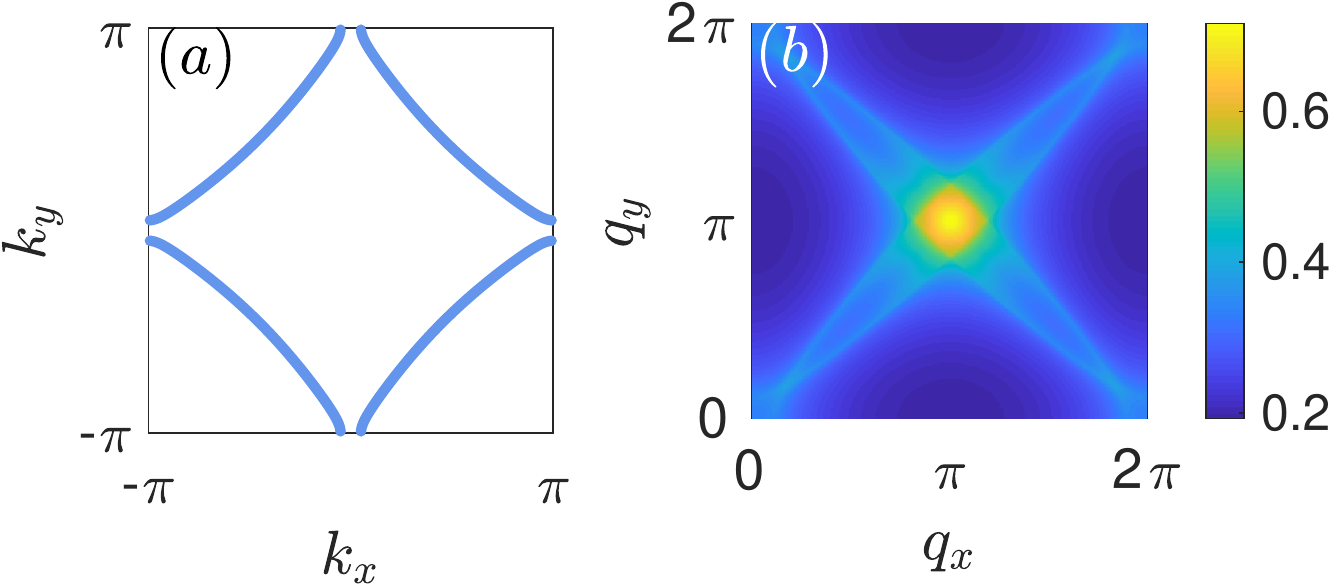}
\caption{(a) Fermi surface and (b) real part of bare spin susceptibility \mbox{$\chi_0(\qv,\omega=0)$} (b) for electron filling $n=0.90$ and $t^\prime=-0.15$.}
       \label{fig:FSandchi0}
\end{figure}

We calculate the spin-singlet $(s)$ (spin-triplet $(t)$) gaps by symmetrizing (antisymmetrizing) the effective interactions, $V^{s/t}(\kv,\kv') =\frac{1}{2}[V(\kv,\kv')\pm V(-\kv,\kv')]$.    
The superconducting gap equation 
\begin{align}
\Delta_\kv=-\sum_{k'}V^{s/t}(\kv,\kv')\frac{\Delta_\kv}{2E_\kv}\tanh\Big(\frac{\beta E_\kv}{2}\Big),
\end{align}
with $E_\kv=\sqrt{\xi_\kv^2+| \Delta_\kv|^2}$ is linearized by setting $E_\kv=|\xi_\kv| $. This gives 
the leading and subleading superconducting instabilities at $T_c$ and amounts to a calculation of the eigenvalues $\lambda_i$ and corresponding eigenvectors $g_i(\kv)$ of the matrix
\begin{eqnarray}
  M^{s/t}_{\kv,\kv'}=-\frac{1}{(2\pi)^2}\frac{l_{\kv'}}{v_F(\kv')} V^{s/t}(\kv,\kv'),
  \label{eq:LGEmatrix}
\end{eqnarray}
with $\kv,\kv'$ restricted to the Fermi surface. Here $l_{\kv'}$ is the length of the Fermi surface segment, $v_F(\kv')$ is the Fermi velocity at $\kv'$. The largest eigenvalue corresponds to the leading instability, but additional details of the pairing structure are reflected in the subleading solutions. Below, we compare the solutions of the linearized gap equation to the obtained instabilities from Quantum Monte Carlo DCA.

\begin{table}[tb]
\centering
\begin{tabular}{ k r r r d d d d }
\hline
\mc{1}{ KL} &\mc{3}{ RPA} &\mc{4}{ DCA} \\
\hline
\hline 
$U^2\chi$ & U=0.1 &U=1 & U=1.3  & U=2 & U=4 & U=6 & U=8 \\
\hline 
\textcolor{d4blue}{$d(4)$}&  \textcolor{d4blue}{$d(4)$} & \textcolor{d4blue}{$d(4)$} &{\textcolor{d4blue}{$d(4)$}} & \textcolor{d4blue}{$d(4)$} &  \textcolor{d4blue}{$d(4)$}     & {\textcolor{d4blue}{$d(4)$}} & { \textcolor{d4blue}{$d(4)$}} \\
\textcolor{ppurple}{$p^\prime(6)$}&\textcolor{ppurple}{$p^\prime(6)$} &\textcolor{d12orange}{$d(12)$} &\textcolor{g8yellow}{$g(8)$}& \textcolor{d12orange}{$d(12)$} & \textcolor{d12orange}{$d(12)$} & \textcolor{d12orange}{$d(12)$} & \textcolor{d12orange}{$d(12)$} \\
\textcolor{d12orange}{$d(12)$}&\textcolor{d12orange}{$d(12)$} &\textcolor{g8yellow}{$g(8)$}&\textcolor{d12orange}{$d(12)$} &\textcolor{g8yellow}{$g(8)$}& \textcolor{g8yellow}{$g(8)$}     & \textcolor{g8yellow}{$g(8)$} & \textcolor{g8yellow}{$g(8)$} \\
\textcolor{g8yellow}{$g(8)$}&\textcolor{g8yellow}{$g(8)$}&  \textcolor{ppurple}{$p^\prime(6)$}& \textcolor{ppurple}{$p^\prime(6)$}  &$d (20)$ & $d(4)$      & $d(12)$ & $d(12)$ \\
\mc{4}{}&\textcolor{ppurple}{$p^\prime(6)$}&  \textcolor{ppurple}{$p^\prime(6)$}   & $s^\prime(8)$ & $s^\prime(8)$\\
\mc{6}{}  & $d(12)$ & $d(12)$ \\
\mc{6}{}& $s^\prime(8)$ & $s^\prime$ \\
\mc{6}{}    & $d(12)$ & $s^\prime$ \\
\mc{6}{}    & \textcolor{ppurple}{$p^\prime(6)$} & \textcolor{ppurple}{$p^\prime(6)$} \\
\hline
\end{tabular}
\caption{The leading superconducting instabilities of Kohn-Luttinger (KL), RPA and DCA calculations for  $n=0.90$ and $t'=-0.15$ as a function of $U$. (For the $U=8$ case, $t'=0$). The gap symmetry is stated by a letter and the number of nodes at the Fermi surface in parenthesis. The following structures appear: $A_{1g} ~[\cos(k_x)+\cos(k_y)$ denoted by $s^\prime(8)]$, $A_{2g} ~[\sin k_x \sin k_y(\cos(k_x)-\cos(k_y))$ denoted by $g(8)]$, $B_{1g} ~[\cos(k_x)-\cos(k_y)$ denoted by $d(4) $ and higher order $(\cos(2k_x)+\cos(2k_y))(\cos(k_x)-\cos(k_y))$ denoted by d(12)$]$ and the $E_u ~[ (\cos(k_x)-\cos(k_y))\sin(k_x)$  denoted by $p^\prime(6)$, but with nodes displaced slightly away from the zone diagonal$]$.
For simplicity we only state instabilities that appear before the leading triplet solution (with the exception of $U\leq 0.1$). In the last $s^\prime$ solutions of the $U=8$ column, the number of nodes is undecided due to system size limitations.}
\end{table}

\section{Results}
\begin{figure}[t]
\centering
  	  \includegraphics[angle=0,width=\linewidth]{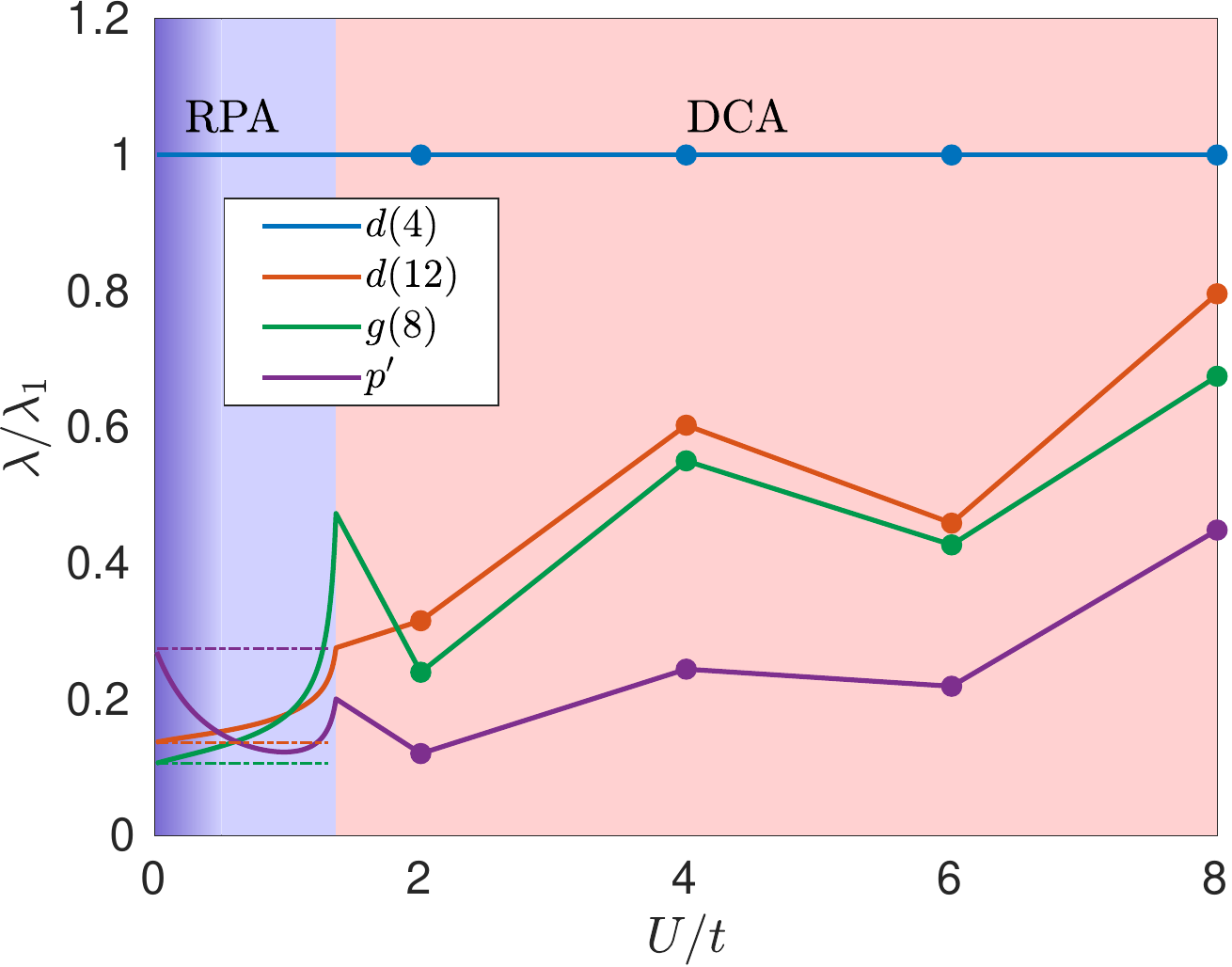}
  	  \caption{Relative eigenvalues $\lambda_i/\lambda_1$ for subleading instabilities $i\in\{d(12),g(8),p^\prime\}$ as a function of interaction $U$ at density $n=0.90$. The eigenvalues are derived from Kohn-Luttinger (dashed-dotted lines) and RPA (full lines) in the regime $U=0-1.36$ (blue area) and from DCA for $U=2,4,6,8$ (pink area). $\lambda_1$ refers to the leading eigenvalue, which always belongs to the $d(4)$ solution. For clarity we have omitted subleading singlet instabilities appearing between $g(8)$ and $p^\prime$ for $U\geq 2$.}
  	  \label{fig:Solutions_vs_U}
\end{figure}

We focus first on a case near half-filling with electron density $n=0.90$ and nearest-neighbor hopping $t'=-0.15$ and explore the role of increasing $U$ from weak to strong coupling. The associated non-interacting Fermi surface and bare susceptibility $\chi_0(\qv,\omega=0)$ featuring pronounced $(\pi,\pi)$-centered fluctuations are shown in Fig.~\ref{fig:FSandchi0}. For this band, the solution to the DCA gap equation, Eq.~(\ref{eq:DCAgapeqn}), is given by a $d_{x^2-y^2}$ symmetric gap function with four nodes as the leading instability for all values of coupling strength $U=2,4,6,8$. (For $U=8$ we set $t'=0$ to avoid the sign problem.) The evolution of the leading and subleading DCA instabilities as a function of $U$ is shown in Table I and plotted in Fig.~\ref{fig:Solutions_vs_U}. Here, we limit the discussion to even-frequency solutions, but note that subleading odd-frequency solutions also exist. In contrast to DCA, both the KL and RPA schemes are limited to small values of the Coulomb interaction of $U = \mathcal{O}(t)$, and RPA is additionally sensitive to the inherent magnetic instability that occurs upon increasing $U$. As seen from Table I and Fig.~\ref{fig:Solutions_vs_U}, in all the DCA cases the leading solution is the lowest order $d_{x^2-y^2}$ solution with four nodes along the Brillouin zone diagonals. Upon increasing $U$, the subleading DCA instabilities approach the leading $d(4)$ instability as inferred from Fig.~\ref{fig:Solutions_vs_U}, and additional nodal singlet solutions appear in-between the $d_{x^2-y^2}$ state and the highest triplet state denoted $p^\prime$ in Table I and Fig.~\ref{fig:Solutions_vs_U}. The number of gap nodes resolved at the Fermi level is sensitive to the cluster size; at $U=2$ which is calculated for a cluster size of $N=64$, twenty nodes are resolved at the Fermi surface for the third subleading $d_{x^2-y^2}$ DCA solution. In comparison, the third subleading solution at $U=4$ exhibits only four nodes. While this could be a real effect due to the increased interaction strength it may also simply be due to the smaller cluster size of $N=32$. For $U=6$ a larger number of subleading solutions appear, many of the same nodal structure, e.g. the $d_{x^2-y^2}$ solutions with twelve nodes at the Fermi surface, denoted $d$(12), which are distinguished by a change of spectral gap weight at different parts of the Fermi surface. 

Turning next to the KL and RPA results for the same band, the hierarchy of the leading pairing solutions are displayed also in Table I and Fig.~\ref{fig:Solutions_vs_U}. As expected for this filling, both methods predict a leading  $d_{x^2-y^2}$ state and agree on the hierarchy of the subleading solutions at the lowest $U$. In contrast to DCA, however, in the low-$U$ limit, the triplet solution denoted $p^\prime$ becomes the second leading instability. This is due to the proximity to the van Hove singularity, which is known to enhance the effective triplet pairing interaction~\cite{Romer2015}. For $t'=-0.15$ the critical density for which the van Hove saddle points reside at the Fermi surface is $n_{\text{van Hove}}=0.875$ and at $n=0.90$ we are thus not far from this regime.

\begin{figure*}[t]
\centering
  	  \includegraphics[angle=0,width=0.3\linewidth]{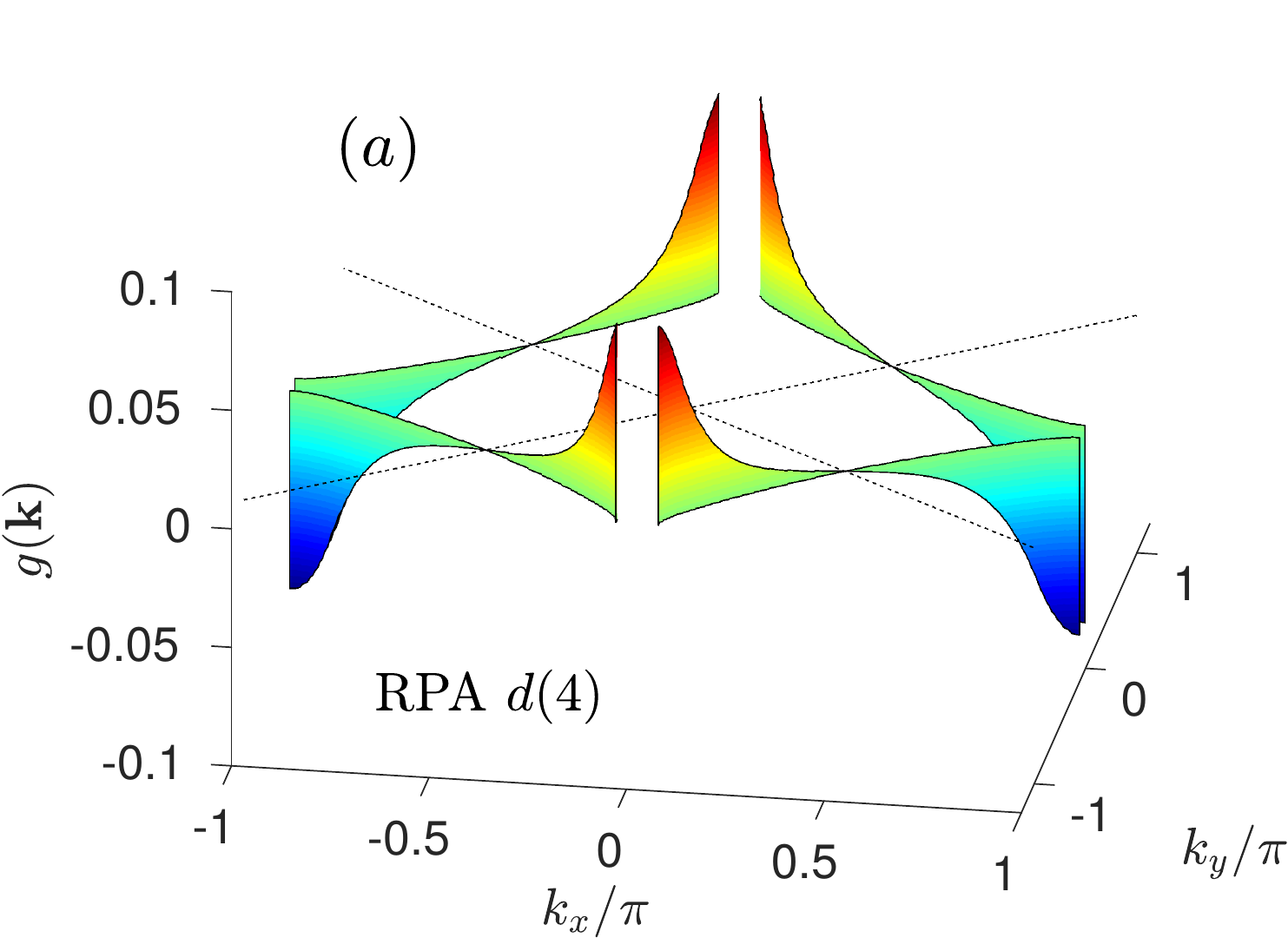}
  	  \includegraphics[angle=0,width=0.3\linewidth]{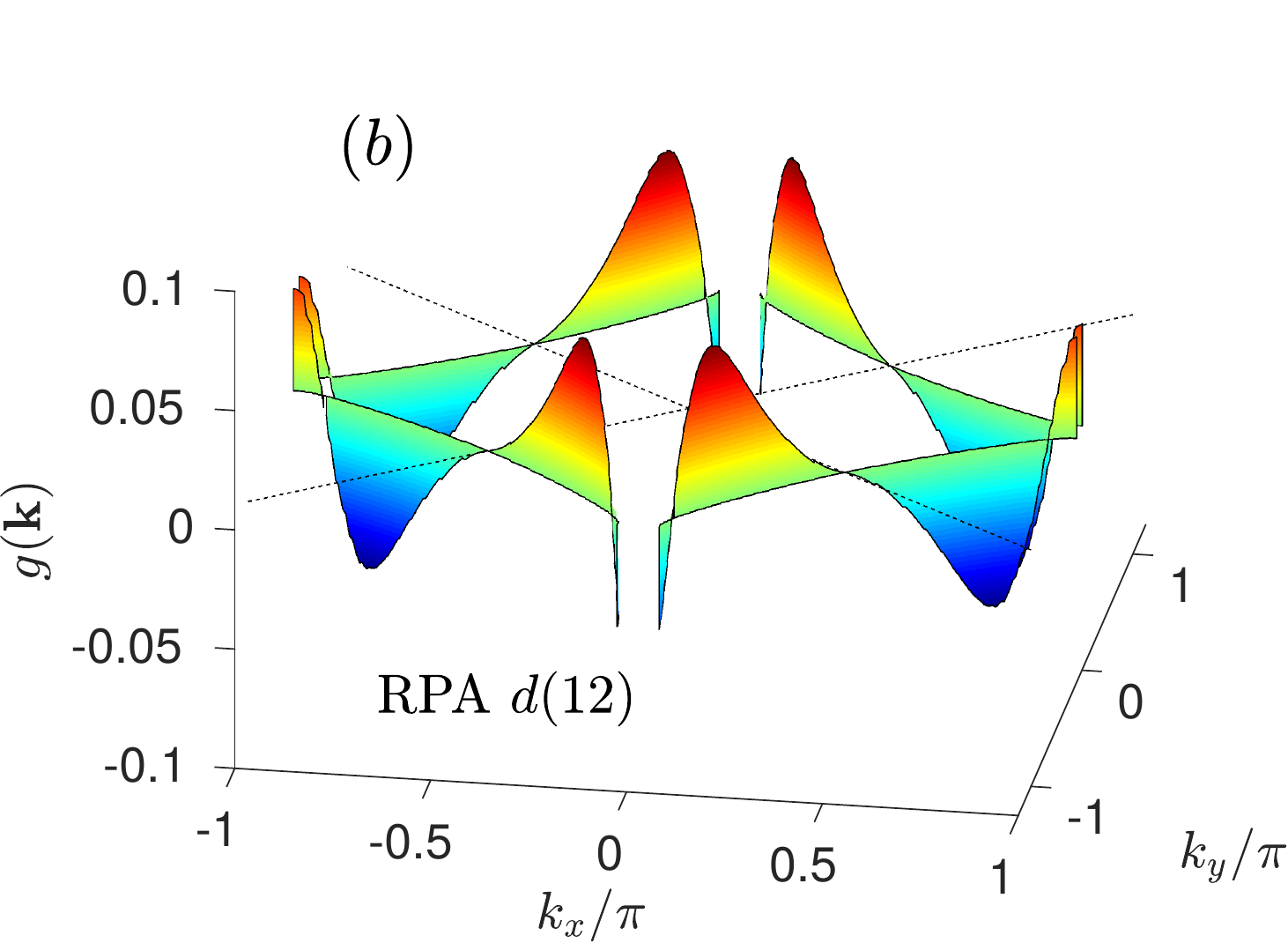}
  	  \includegraphics[angle=0,width=0.3\linewidth]{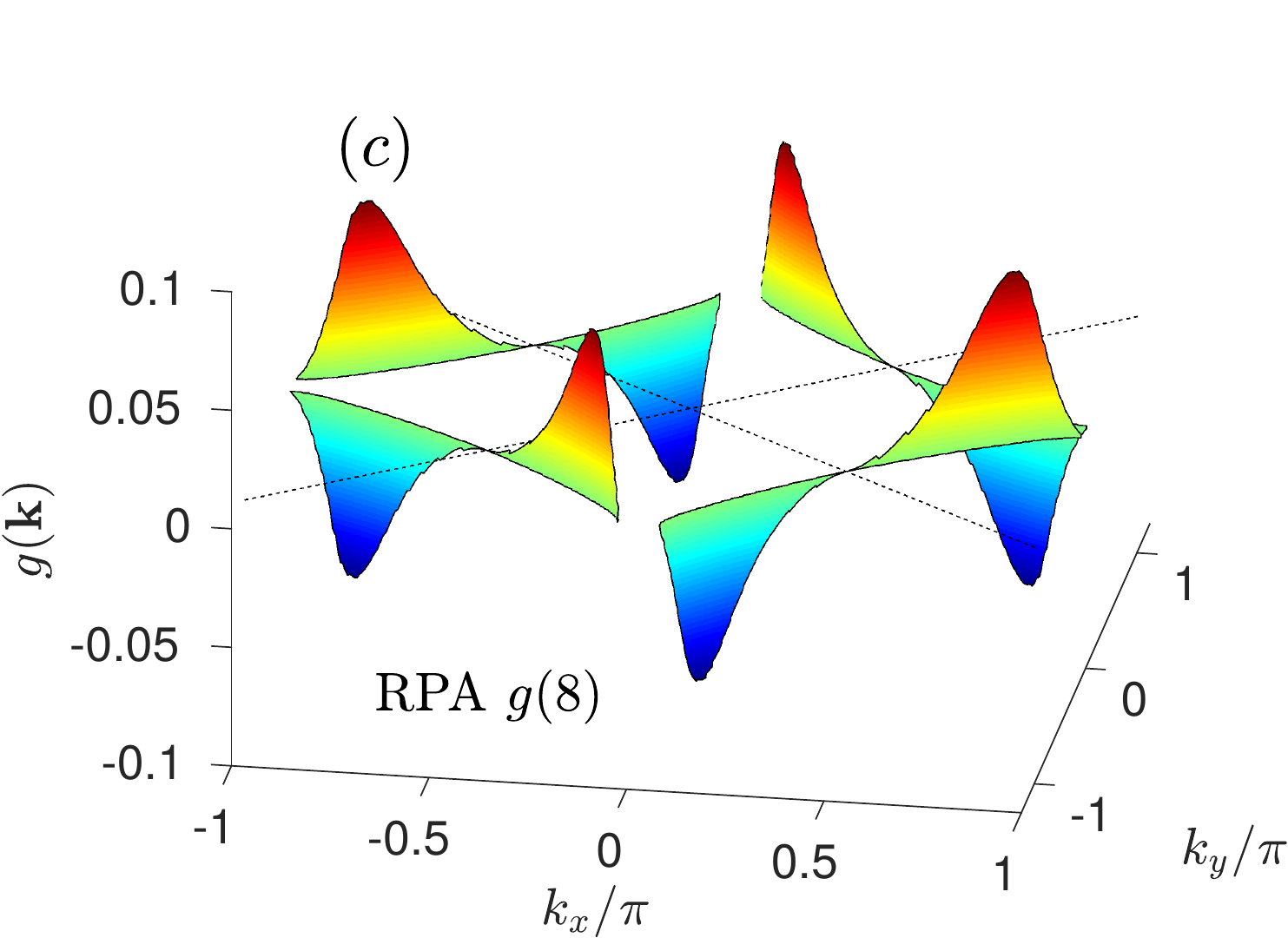}  	  
\includegraphics[width=0.3\linewidth]{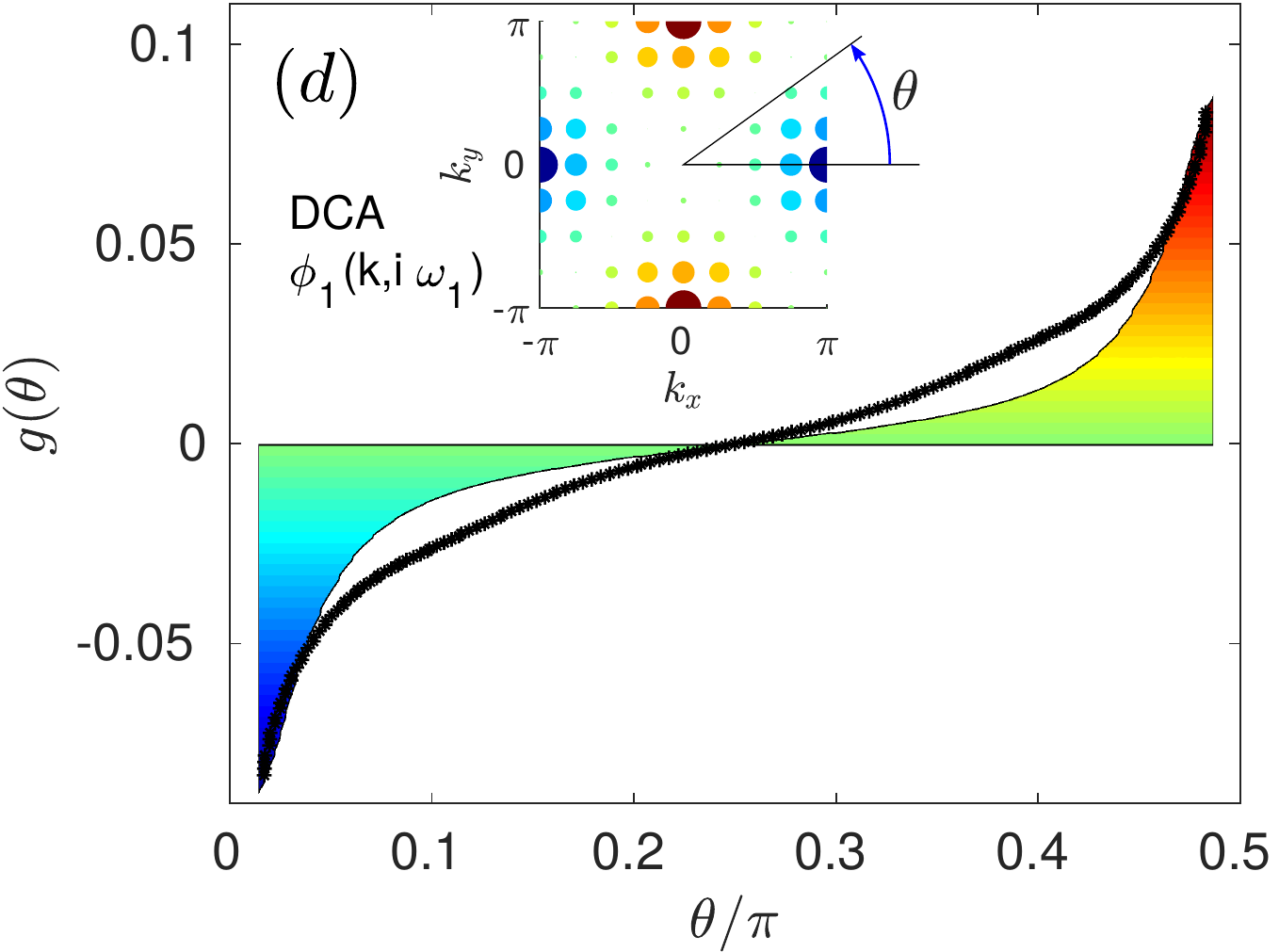}
  \includegraphics[width=0.3\linewidth]{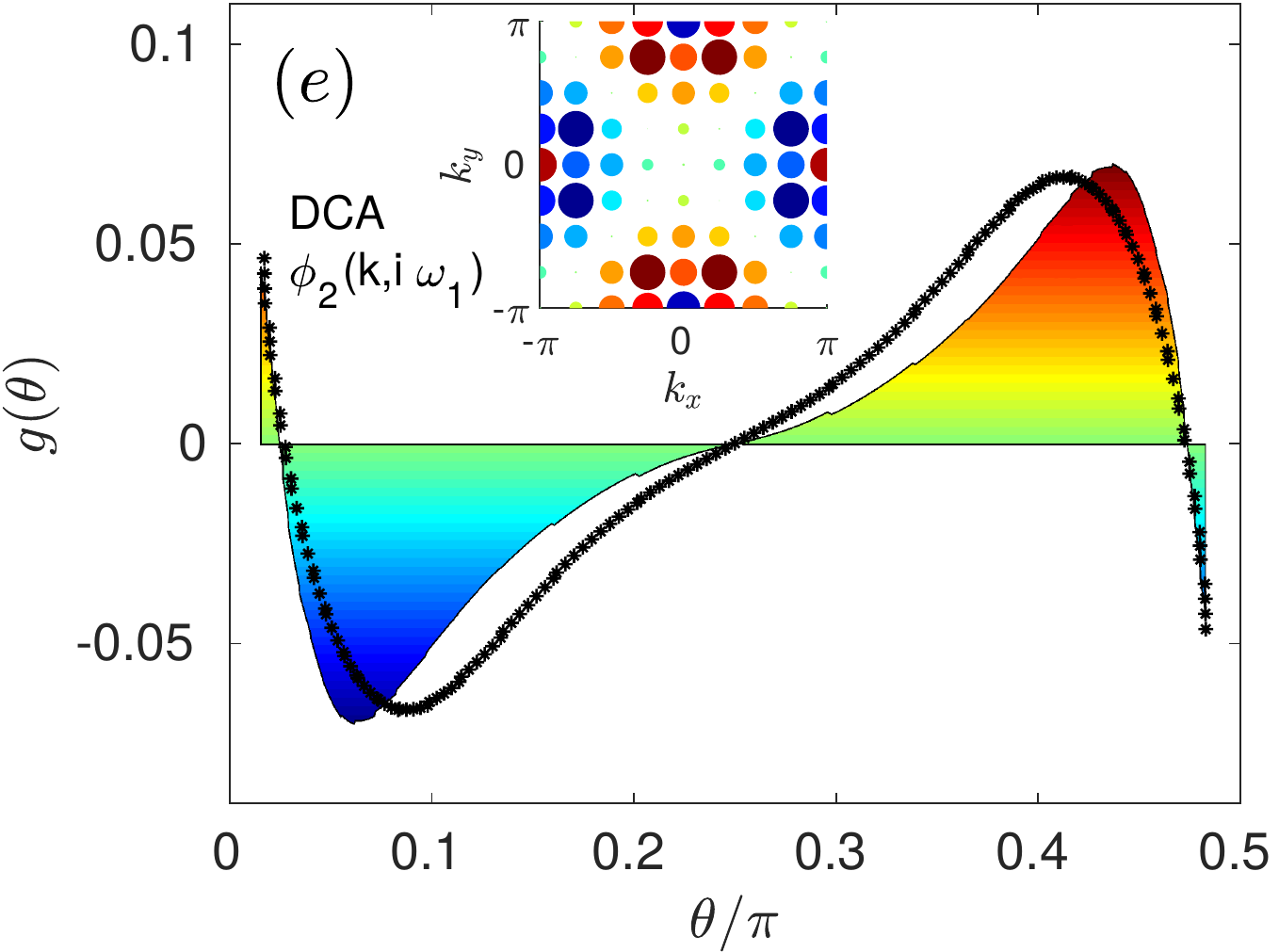}
 \includegraphics[width=0.3\linewidth]{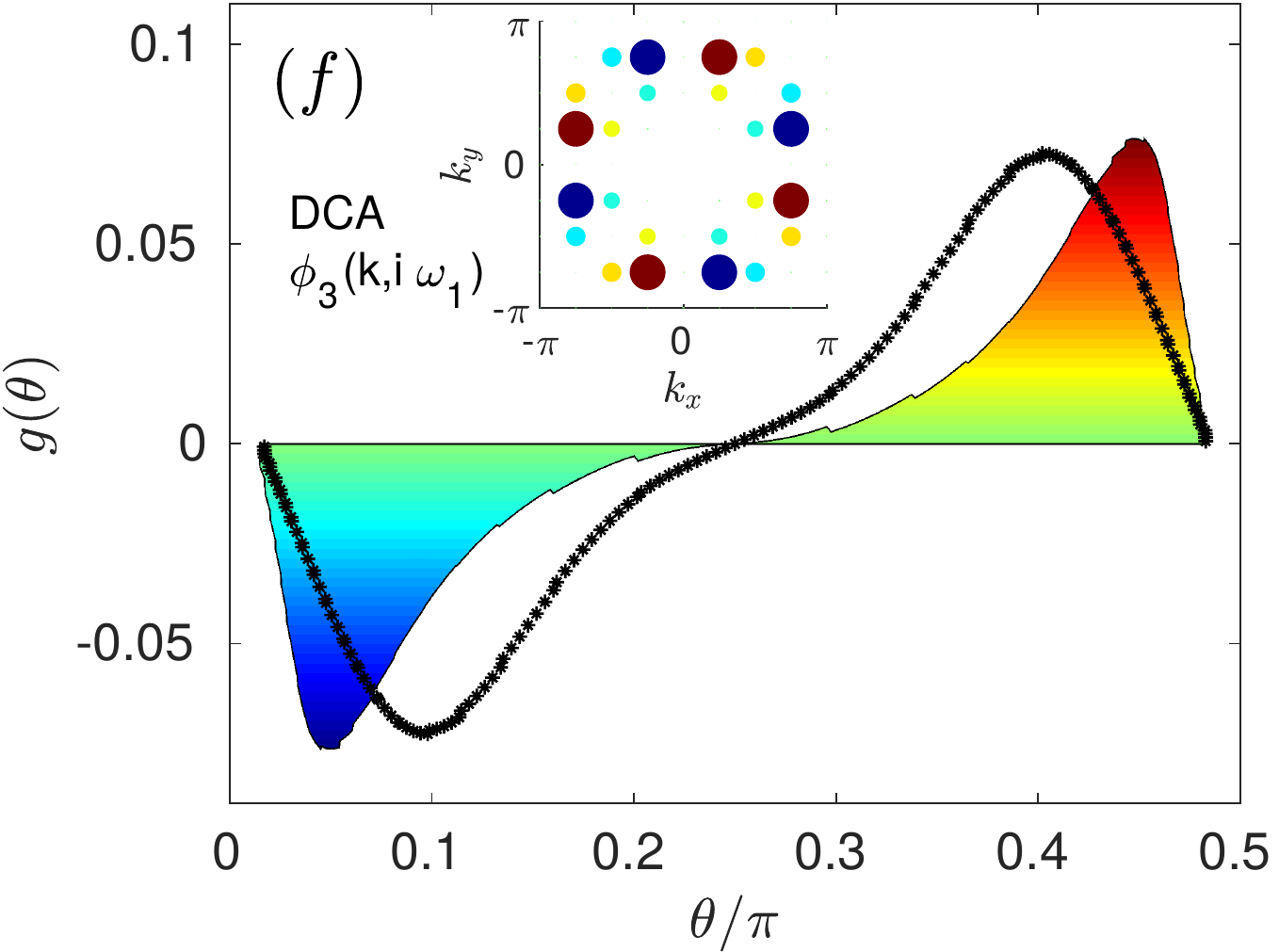}
  	   \caption{(a-c) Three leading solutions of the RPA model ($U=1$) with the leading $d_{x^2-y^2}$ solution, the sub-leading higher order $d_{x^2-y^2}$ solution with 12 nodes and third leading  $g$-wave solution. (d-f)  The results of the DCA calculation, $\phi_\alpha({\bf K},i\omega_1)$ at the lowest frequency are shown as insets. Comparison of RPA and DCA results by angle dependent gap plots in the first quadrant of the Brillouin zone. The angle $\theta$ is defined in the inset in $(d)$. The three leading solutions of the RPA calculation ($U=1$) are shown in color and the DCA results ($U=2$) are shown by black symbols. The DCA results are interpolated to the Fermi surface of the non-interacting system to allow a direct comparison although the DCA calculation is only done for $8\times 8$ discrete values.}
  	   
       \label{fig:RPA_BS_singletsolutions}
\end{figure*}

Upon increasing $U$ (but still within the RPA regime), the $p^\prime$ solution rapidly drops and, as seen from Table I and Fig.~\ref{fig:Solutions_vs_U}, there is excellent agreement with the DCA pairing hierarchy near the regime of $U=1$. At larger $U$, still within the RPA calculation, the magnetic instability is approached, and the two nearly degenerate instabilities $d(12)$ and $g(8)$ are interchanged. This is an artifact of the RPA approach which can be understood in the following way.
The $g(8)$ solution increases more steeply because it takes full advantage of the strongly enhanced susceptibility.  
Unlike $d(12)$, the $g(8)$ solution does not have any nodes at the Fermi surface segments in the large gap regions close to $(\pi,0)$ and symmetry-related points. The symmetry-imposed nodes of the $g(8)$ solution along the zone axes do not inhibit gap formation, since the Fermi surface segments do not close at $(\pi,0)$ and symmetry-related points. Nevertheless, except from such caveats as (over)sensitivity to band details or magnetic instabilities, the overall evolution of the leading superconducting solutions as discussed here highlights the agreement of the methods, and points to an adiabatic continuity between weak- and strong-coupling pairing solutions of the Hubbard model near the half-filled regime.

Next, we compare the detailed properties of the gap solutions obtained by DCA to the results of RPA (at $U=1$). In Fig.~\ref{fig:RPA_BS_singletsolutions}, the three leading instabilities of both approaches are displayed. As seen, there is remarkable agreement between the two methods, giving in both cases a leading $d_{x^2-y^2}$ solution with four diagonal nodes, i.e. $\Delta_d(\kv)=\frac{\Delta}{2}[\cos(k_x)-\cos(k_y)]$, but with strong gap enhancements around $(0,\pm\pi)$ and $(\pm \pi,0)$. In RPA this enhancement is caused by the large density of states present in those regions of $k$-space due to the proximity of the van Hove singularity. A similar effect at different doping levels was discussed in Ref.~\onlinecite{Romer2015}. In the linearized gap equation this effect enters via the inverse of the Fermi velocity in the matrix elements of Eq.~(\ref{eq:LGEmatrix}) as well as in the amplification of the bare spin susceptibility. However, the enhancement is also found in the DCA approach where we interpolate the solution to the Fermi surface of the non-interacting system, see from Fig.~\ref{fig:RPA_BS_singletsolutions}(d). For $U=2$ we expect the Fermi surface to be very similar to that of the non-interacting system. From Fig.~\ref{fig:RPA_BS_singletsolutions}(d) we see that the gap enhancements are robust towards the inclusion of self-energy effects. 

The two sub-leading solutions shown in Fig.~\ref{fig:RPA_BS_singletsolutions} consist of a $d_{x^2-y^2}$-wave solution with twelve nodes $d(12)$, and a lowest order $g$-wave state with eight nodes $g(8)$. These solutions are close in energy and are both strongly suppressed compared to the leading $d_{x^2-y^2}$ gap with four nodes. For the second and third leading solutions, the RPA approach produces strong gap enhancements at the Fermi surface points closest to $(0,\pm\pi)$ and $(\pm \pi,0)$ whereas this effect is less pronounced in the DCA calculations, especially for the third leading $g$-wave solution. This may arise from the fact that the $g$-wave solution has nodes along the zone axes and DCA, with fewer $\kv$-points to sample the Brillouin zone as shown in the inset of Fig.~\ref{fig:RPA_BS_singletsolutions}(f), therefore does not capture the enhancement effect for this solution.

\begin{figure*}[tb]
\centering
  	  \includegraphics[angle=0,width=0.495\linewidth]{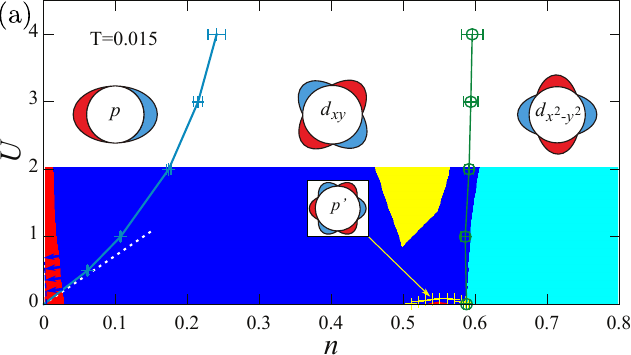}
  	    	  \includegraphics[angle=0,width=0.495\linewidth]{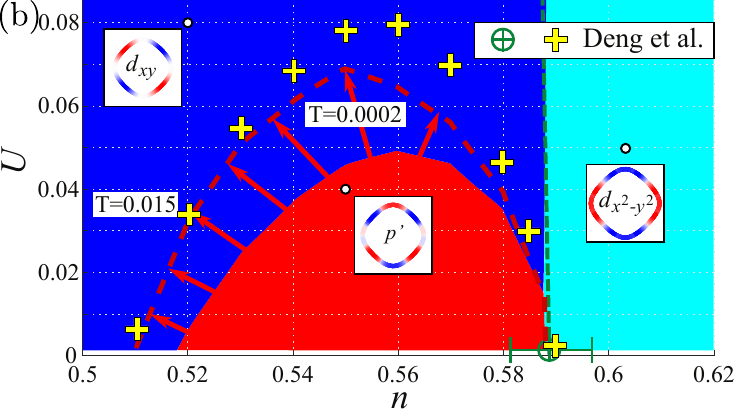}
       \caption{(a) Symmetry of the leading superconducting instability as a function of $U$ and filling $n$ with $t'=0$ and $T=0.015$. RPA results are indicated by filled colors superimposed in the ground state phase diagram obtained in Ref.~\cite{deng15} shown by open symbols and lines. The blue line displays the phase boundary between a simple harmonic $p$-wave ($\sin(k_x)/\sin(k_y)$) and $d_{xy}$, and the green line indicates the phase boundary between $d_{xy}$ and $d_{x^2-y^2}$. The yellow region indicates an $s'$ phase\cite{Kreisel2017_JSNM}. A region of higher order $p'$-wave (sketch as inset \cite{deng15}) is visible in both approaches for small values of $U$ around a filling of $n=0.55$. The triplet $p'$ state features six or ten nodes depending on temperature. Lastly, the region of leading triplet instability at very low filling is sensitive to resolution and temperature; it essentially vanishes as $T$ is decreased to $T=0.0002$ (small blue arrows and blue dashed line almost coinciding with the $y$-axis). (b) Zoom in of the phase diagram close to $n =0.55$ and comparison to Ref.~\cite{deng15}. The insets show solutions $g({\bf k})$ as obtained from the LGE at the parameters marked by white points. The dashed red line (and the small red arrows) shows how the boundary between the $p^\prime$ and $d_{xy}$ changes within the LGE when $T$ is lowered from $T=0.015$ to $T=0.0002$.}
       \label{fig:DengComparison}
\end{figure*}

Encouraged by the overall agreement between KL/RPA and DCA, we next compare the RPA results obtained here with previous reports in the literature. In Fig.~\ref{fig:DengComparison}, we show a direct comparison of the ground state phase diagram of Ref.~\onlinecite{deng15} obtained from diagrammatic Monte Carlo simulations and our RPA calculations. As seen, there is qualitative and in most cases nearly quantitative agreement between the different methods. For example, upon hole-doping the instability from the $d_{x^2-y^2}$ state to the $d_{xy}$ state occurs almost simultaneously in the two methods. Furthermore, both KL/RPA and diagrammatic Monte Carlo simulations find a $p'$-wave triplet state to become leading for small values of $U$ around a filling of $n=0.55$. 

Most often, triplet solutions become favorable when the system approaches a van Hove singularity regime~\cite{Romer2015,Simkovic_2016,greco19}. To second order in $U$, the pairing vertices are given by
\begin{align}
 V_{\rm sing}^{\rm RPA}&=U+\frac{U^2}{2}[\chi_0(\kv-\kv')+\chi_0(\kv+\kv')]+\mathcal{O}(U^3), \\
 V_{\rm trip}^{\rm RPA}&=\frac{U^2}{2}[-\chi_0(\kv-\kv')+\chi_0(\kv+\kv')]+\mathcal{O}(U^4).
\end{align}
In the absence of a $\qv=0$ peak structure in the susceptibility, the triplet pairing cannot take advantage of the attractive contribution to the  pairing kernel $-\frac{U^2}{2}\chi_0(\kv-\kv')$. Usually, such a peak is what renders the triplet solution favorable in the vicinity of a van Hove instability. However, the region of triplet superconductivity evident from Fig.~\ref{fig:DengComparison}(a) around  $n=0.55$ at the smallest $U$ has a different origin (at $t'=0$ the van Hove singularity occurs at the Fermi level for filling $n=1$). 

Around $n=0.55$ the system is in an interesting cross-over regime, where the spin susceptibility shows prominent features at nesting vectors $\Qv \simeq (\pi,\pm\frac{\pi}{2})/(\pm\frac{\pi}{2},\pi)$, which lie right in-between nesting vectors along the zone diagonal and zone axes, driving the singlet $d_{x^2-y^2}$ and $d_{xy}$ solutions, respectively. 
The odd parity $p^\prime$ solution most optimally accommodates this nesting structure, but since it is not supported by a $\qv=0$ peak, it is rather fragile and becomes rapidly suppressed as $U$ increases.
The latter can be understood from the fact that the spin susceptibility exhibits extended ridge-like structures which are most pronounced around $(\pm\pi,0)/(0,\pm\pi)$. Upon increasing $U$, these ridges will dominate the effective pairing and drive the system into the singlet $d_{xy}$ solution.

In Fig.~\ref{fig:DengComparison}(b) we show a zoom-in of the phase diagram relevant to the triplet $p^\prime$ phase. As seen, the phase boundary of the $p^\prime$ phase to the $d$-wave states agrees remarkably well with the diagrammatic Monte Carlo calculations by Deng {\it et al.}\cite{deng15} at the lowest temperatures. The detailed gap structure of the superconducting $p^\prime$ state features ten nodes as shown by the inset in Fig.~\ref{fig:DengComparison}(b). This is slightly different from the illustration shown in Fig.~\ref{fig:DengComparison}(a) from Ref.~\onlinecite{deng15}, but similar to the gap structures discussed in Ref.~\onlinecite{Simkovic_2016}.

We end with a brief discussion of the pairing instabilities in the low-density regime $n\lesssim 0.3$. As seen explicitly in Fig.~\ref{fig:DengComparison}(a), the low-density limit hosts a triplet $p$-wave superconducting phase. We stress that this is a standard two-node $\sin(k_x)/\sin(k_y)$ $p$-wave state distinct from the $p^\prime$ triplet state discussed above. The possibility of a transition from $d_{x^2-y^2}$ to $d_{xy}$ or $p$-wave superconductivity in the low-density regime of the weakly repulsive 2D Hubbard model was discussed early on by Baranov and Kagan~\cite{Baranov1992}, and by Chubukov and Lu~\cite{Chubukov92} who analyzed the behavior of the pairing vertex in symmetry-distinct pairing channels as a function of band parameters. The more recent diagrammatic Monte Carlo calculations by Deng {\it et al.},\cite{deng15} mapped out the phase boundaries between the $p$-, $d_{xy}$-, and $d_{x^2-y^2}$-wave pairing solutions in the low-density and low-$U$ limits, reproduced in Fig.~\ref{fig:DengComparison}. As seen from Fig.~\ref{fig:DengComparison}, even though there is substantial overall agreement to the RPA results, the low-density regime stands out as exceptional in this comparison between the methods. At the lowest $T$ and in the limit $n, U\rightarrow 0$ the preferred state from the RPA study is $d$-wave with near degeneracy between $d_{xy}$-, and $d_{x^2-y^2}$-wave pairing. At larger $U$, however, as seen from Fig.~\ref{fig:DengComparison}, RPA does not capture the preference for $p$-wave pairing in the dilute limit. This result, however, is not surprising since the lowest order diagrams included in the RPA procedure are known to not capture the tendency for $p$-wave pairing in the dilute limit. Only by including higher order vertex renormalizations does $p$-wave pairing get supported. This was shown initially by Chubukov who analyzed the third order diagrams for renormalization of the fermionic scattering amplitude in 2D, and found that the vertex renormalization in the particle-particle channel is crucial for realizing the $p$-wave state at low densities\cite{Chubukov93}. Subsequent studies confirmed the importance of $\mathcal{O}(U^3)$ vertex corrections for stabilization of $p$-wave pairing at low density\cite{Fukazawa,Takahashi1999}. As a consistency check we applied the DCA machinery to calculate the leading instability at $U=4$, $t^\prime=0$ and $T=0.0125$ at fillings $n=0.15$ and $n=0.20$ (due to resolution we cannot address lower $n$ by this method). In the first case ($n=0.15$) we obtained indeed a leading $p$-wave solution even for $U=4$, while $d_{xy}$ is the preferred state at $n=0.20$. This points to a rough agreement with the phase boundary obtained by diagrammatic Monte Carlo simulations in Ref.~\cite{deng15}, and again suggests a smooth crossover of superconductivity from weak to strong interactions.

\section{Conclusions}

While there is general agreement that the leading  Cooper pairing instability  of the Hubbard model close to half-filling is the $d_{x^2-y^2}$-wave state, and work on the $t-J$ model valid in this regime corresponding to very large $U$ suggests the same, rather less is known consensually about the rest of the Hubbard model pairing phase diagram, including fillings far from $n=1$ and intermediate to strong $U$.  These regimes are not simply of academic interest, but may well represent reasonable descriptions of a variety of unconventional superconductors, including cuprates, organic Bechgaard salts, heavy fermion materials, iron-based superconductors, and ultra-cold fermionic gasses.  In this work, we have compared different approximate methods, expected to be valid in different correlation regimes, to predict the leading and subleading superconducting instabilities in these less-studied situations.  We find that spin and charge fluctuation exchange pairing calculated from both KL (small $U$) and RPA methods, and a DCA Quantum Monte Carlo approach (intermediate to strong $U$) compare rather favorably to each other, suggesting a smooth crossover in pairing states within the Hubbard model from weak to strong coupling at all fillings. Our results compare well  to recent diagrammatic Monte Carlo calculations, with the exception of the regime of very small electron density where weak-coupling approaches need to be cured by vertex corrections. The agreement with RPA allows for a transparent explanation of the physics of several of these less well-known pairing phases. Clearly the hypothesis of adiabatic connectivity of pair states from weak to strong coupling needs further scrutiny and investigations by other methods capable of handling electron pairing in the strongly correlated regime. 

\section{Acknowledgements}

A.T.R.\ and B.M.A.\ acknowledge support from Lundbeckfond fellowship (grant A9318), and the Carlsberg Foundation. P.J.H. was supported by the U.S. Department of Energy under Grant No. DE-FG02-05ER46236. The DCA calculations (T.A.M) were supported by the Scientific Discovery through Advanced Computing (SciDAC) program funded by U.S. Department of Energy, Office of Science, Advanced Scientific Computing Research and Basic Energy Sciences, Division of Materials Sciences and Engineering.

%

\end{document}